\documentclass[12pt]{article}
\usepackage[dvips]{graphicx}
\usepackage{floatflt}
\usepackage{color}
\usepackage{colortbl}

\newcommand{\be}{\begin{equation}}
\newcommand{\ee}{\end{equation}}
\newcommand{\bean}{\begin{eqnarray}}
\newcommand{\eean}{\end{eqnarray}}
\newcommand{\bea}{\begin{eqnarray*}}
\newcommand{\eea}{\end{eqnarray*}}

\newcommand{\bc}{\begin{center}}
\newcommand{\ec}{\end{center}}
\newcommand{\q}{\quad}
\newcommand{\ds}{\displaystyle}

\suppressfloats[!]

\begin{document}

\begin{center}
\Large{\bfseries On the predictions of dip-effect in $Q^2$ dependence of 
                 electromagnetic and electroweak formfactors of $\pi$-meson 
                 decays  and their experimental verification.}

\vskip 5mm

N.B.~Skachkov

\vskip 5mm

{\small
{\it
Joint Institute for Nuclear Reseach,Laboratory of Nuclear Problems,\\
JINR, Dubna, 141980, RUSSIA
}
\\
$\dag$ {\it
E-mail: skachkov@cv.jinr.ru}
}
\end{center}

\vskip 5mm

\begin{center}
\begin{minipage}{150mm}
\centerline{\bf Abstract}
         
    In present note the arguments in favour of high statisics
 measurements of the formfactors that describe  $\pi$-meson 
 decays into a lepton-antilepton pair plus a photon are given.
 It is shown that these formfactors may contain an important
 information on the dynamics of quark motion  inside a hadron.



\end{minipage}
\end{center}

\vskip 5mm
   
\section{   Introduction.}

     The aim of the present note is to discuss the arguments in
   favour  of future high statistic measurements of $\pi$-mesons
   decay formfactors that contain the information on quark structure
   of mesons. The decays of light $\pi$-meson such as
   $\pi^0 \to \gamma + e^+e^-$  and   $\pi^{\pm} \to \gamma + e^\pm \nu$ 
   (in what follows they would be denoted as $\pi^0 \to \gamma+ee$
   and $\pi\to \gamma+e\nu$ processes) are of a big interest from the 
   view point of studying  the dynamics of quark motion  inside a 
   hadron. The pioneering works on kinematics of these processes and the 
   relation between the vector $F_V$ and the axial-vector $F_{A}$ 
   formfactors as well as on the relation with the width of 
   $\pi^0 \to \gamma\gamma$  process and also on the account of 
   Inner Bresstrahlung amplitude conrtibution one can find in \cite{hist}.
  
     There is a plenty of different models based on current algebra, 
   vector dominance model, chiral field theories and etc. (see, for 
   instance, the reviews \cite{PhysRep}) 
   that were  proposed to describe the formfactors dependence 
   on a square of the  invariant mass of a final state lepton 
   pair. This variable can also be interpreted as a square of a 
   4-momentum  $Q^2$ transfered from the hadronic block of the 
   corresponding  Feynman diagram to a leptonic pair. Only 
   one particular prediction on a shape of a 
   $Q^2$- dependence of the decay formfactor, namely, on the
   appearance of a dip in a region of small values of 
   $x=Q^2/M_\pi^2$  variable  \cite{2K3S},  would be 
   discussed here. It should be noted that the high statistic 
   experiment done at Saclay 6 years later  had  presented the 
   data \cite{SACLPL} (for more  details see the reviews in 
   \cite{HFAN} and \cite{HFDIS}) that  may be interpreted as an 
   experimental confirmation of the prediction done in \cite{2K3S}.
   Nevertheless, 
   the systematic errors quoted in \cite{SACLPL} are rather high,
   so  more precise measurements, especially in a region of small 
   values of $x$, would be obviously very important.  
   In this note  it will  be shown also that the analogous dip-effect 
   may  reveal itself  in the formfactors of the electroweak decay
   $\pi^{\pm} \to \gamma + e^\pm \nu$. So, it would be very interesting
   to perform  with good systematic  errors  new high statistic 
   measurements of the behavior of  $F_{\pi^0\to\gamma e^+ e^-} (Q^2)$ 
   formfactor in a region of small values of $Q^2$ as well as to
   collect the analogous data on $F_{\pi^{\pm}\to\gamma e^{\pm}\nu}(Q^2)$
   formfactor.
   
     The prediction done in \cite{2K3S} was obtained in the framework
   of the relativistic  constituent quark model (see references in
   \cite{SKSOL1}-- \cite{SS}))
   which make use  of  a covariant equation for two-body wave function
   \cite{FA} that  was derived on 
   the basis of 3-dimensional quasipotential approach to two-particle
   relativistic equations in  quantum  field theory (QFT) \cite{LTK}.

     The important point of this model is that its mathematical apparatus
   incorporates, in  difference with ordinary QFT  amplitudes, 
   the bound state wave functions of quarks.
   Really, pion is a bound state of  light 
   quark and antiquark  tighten  together by  forces caused by gluon 
   exchange. Therefore, an application of Feynman diagrams (originally 
   proposed in QFT for calculation of scattering amplitudes of 
   particles that are free in an initial state)
   for describing the processes, that include the bound states in initial
   state,  may serve,  definitely, only as some
   perturbative model approximation applied in a region
   where the nonperturbative effects play an essential role. To this
   reason it is natural to expect that the consistent amplitudes to be used
   to describe  the decay processes of mesons have to include the wave 
   functions that take into account the bound state nature  of mesons.   
   
     Finally the prediction of the appearance of a dip is a sequence 
   of three main features of the considered model:

    1. relativistic motion of quarks bounded in spin 0 $\pi$-meson state;

    2. Standard Model (perturbative QCD and SM Feynman diagram technique) form 
       of a  quark propagator that enter the  quarks interaction amplitude
       describing the photon and lepton-antilepton ($e^+e^-$ or $e^\pm \nu$) pair
       production in a final state;

    3. large value of a binding energy in a pion considered as  $q\bar{q}$ bound
       state.     


 
\section{ Main formulas.}

    
     To explain the above statements the main points of the analysis performed 
    in \cite{2K3S} for  $\pi^0 \to \gamma+ee$ case  would be sketched below.
%
%
      The schematic view ( not a Feynman diagram!) of $\pi$-meson decay processes
    that illustrates the corresponding  invariant  amplitudes $M_{\pi\to f}(P|q_1,q_2)$
    at quark level ($k_1$ and $k_2$ are quark and antiquark moments, while $p_l$ and
    $p_{\bar{l}}$ are the moments of the lepton and the antilepton respectively)
    is shown in Fig.~1.

    In Fig.1 $f$ denotes a final state, i.e. $f=\gamma+l\bar{l}$
    for $\pi^0$-decay and $f=\gamma+e\nu$ for $\pi^{\pm}$-decay.
    These amplitudes  may be parameterized  through the  decay
    formfactors  in the following way ($q_{1}=p_{+}+p_{-}=p_{l}+p_{\bar{l}}$):

\begin{center}
\begin{figure}[h]
\vskip-20mm
\hspace*{23mm} \includegraphics[width=15cm,height=10cm]{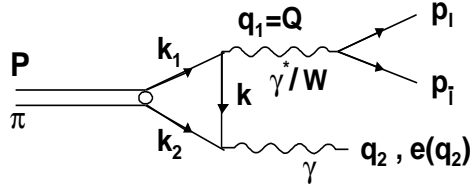}
\vskip-50mm
\caption{\hspace*{0.0cm} The schematic view of $\pi$-meson decay processes .}
\label{fig:1}
\end{figure}
\end{center}

\bean
&&M_{\pi^0\to\gamma e^+ e^-}(q_1,q_2|P)=\frac{F_{\pi^0\to\gamma e^+ e^-}
(q^2_1)}{q_1^2} V_{\mu\nu}(q_2|P) (e)  j_{V}^{\mu}(p_+,p_-) e^{\nu}
(q_2) \, \, ,
\eean
\bean
M_{\pi^{\pm}\to\gamma e^{\pm}\nu}(q_1,q_2|P)&=&\left[F_{V}(q_1^2) \cdot V_
{\mu\nu}(q_2|P)+F_A(q_1^2) \cdot A_{\mu\nu}(q_2|P)\right]\times 
\nonumber
\eean
~\\[-25pt]
\bean
\times \left((e) \frac{G_{F}V_{ud}}{\sqrt{2}}\right) \cdot
j_{V-A}^{\mu}(p_{\pm},p_{\nu}) e^{\nu}(q_2) \, \, .
\eean

    Here  $e^{\nu}(q_2)$ is a polarization 4-vector of a real
    photon with  the 4-momentum $q_2$, ( $e$) is an electric charge
    of a lepton and $G_{F}$  is Fermi coupling constant of weak 
    interaction and $V_{ud}$  ($\sim cos(\theta_c)$)
    is the  element of Cabibbo-Kobayashi-Maskawa matrix. Also   
    the following notations are used for two (``orthogonal'' to 
    each other) structure tensors $V_{\mu\nu}$ and $A_{\mu\nu}$: 
\bean
V_{\mu\nu}(q_1|P)=\epsilon_{\mu\nu\alpha\beta}P^{\alpha}q_1^{\beta} \, \,
\eean
\bean
A_{\mu\nu}(q_1|P)=g_{\mu\nu}(P \cdot q_1) - P_{\mu}q_{1\nu},
\eean
    that define , respectively, the vector and axial formfactors.
    Both electromagnetic (EM) and
    electroweak (EW) final state currents can be defined by one and 
    the same formula:
\bean
j^{\mu}_{V-A}(p_1,p_2)={\bar u}(p_1)\gamma^{\mu}(V-A)u(p_2) \, \, ,
\eean
    with the factor $(V-A)$ defined as 
\bean
V-A=
{\left\{
\begin{array}{ll}
V=1 \, ; A=0  ;\, \q \, ~{\rm for} \, \q \, e^+e^- \, , \\
V=1 \, ; A=\gamma^{5} ;\, \q \, {\rm for} \, \q \, e^{\pm} \nu \, \, .
\end{array}
\right.} 
\eean

    To make clear, why  the same arguments that were used in  \cite{2K3S}
    may be  valid also  for the $F_{V}(q_1^2)$ and $F_{A}(q_1^2)$ 
    formfactors of $\pi^\pm \to \gamma+e\nu$ decay, let us start with
    the definition of the amplitude of the process. 

      An amplitude  $M_{\pi\to f}(q_1,q_2|P)$ of a  $\pi$-meson 
   (considered as a $q\bar{q}$ bound state)
   decay process  may  be presented in a framework of the relativistic
   quark model as a convolution (with the relativistic invariant
   differential volume element of the momentum space 
   $\frac{d^3{\bf k}_1}{2k^0_1}$)  of a covariant 2-body bound 
   state (B)  wave  function $\Psi_{\sigma_1\sigma_2BP}({\bf k}_1)$
   and the final state  interaction  amplitude $T_{q\bar{q}\to f}$,
   taken in a form of Feynman matrix element (see Fig.1). 
 
   Let us consider a general case of ${\pi\to\gamma^*V^*}$ decay,
   where by  $V^*$ the virtual photon $\gamma^*$ or virtual  $W^{*}$ boson
   (see Fig.1) is  denoted. The  transition amplitude is defined, 
   following the guide line of \cite{SS} and \cite{2K3S}, like

   
\bean
&&M_{\pi^0\to\gamma^*V^*}(q_1,q_2|P)=\frac{1}{(2\pi)^{3/2}}\int
\frac{d^3{\bf k}_1}{2k^0_1}
\Psi_{\sigma_1\sigma_2BP}({\bf k}_1)
T^{\sigma_1\sigma_2}_{q\bar q\to\gamma^*V^*}(q_1,q_2;k_1|P) \, \,.
\eean

\vskip 5mm

    In (7) the  summation over quark polarizations $\sigma_{1},\sigma_{2}$ is
    supposed. Now one may  define the formfactor
    of  ${\pi^0\to\gamma^{*}\gamma^{*}}$ transition as follows

\bean
&&M_{\pi^0\to\gamma^*\gamma^*}(q_1,q_2|P)=
F_{\pi^0\to\gamma^*\gamma^*}(q_1^2,q_2^2)
e_1^{\mu} e_2^{\nu} \epsilon_{\mu\nu\rho\sigma}q_1^{\rho}P^{\sigma}=
\nonumber
\eean
\bean
=F_{\pi^0\to\gamma^*\gamma^*}(q_1^2,q_2^2)
e_1^{\mu} e_2^{\nu}V_{\mu\nu}(q_1|P)  \, \, ,
\eean

\vskip 5mm

\noindent
    where $e_1^{\mu}$ and $e_2^{\nu}$ are the  the polarization 4-vectors
    of two virtual photons with the 4-moments (off mass shell)
    $q_1$ and $q_2$ respectively.


    After these definitions one can determine the formfactor
of $\pi^0\to\gamma e^+ e^-$ decay as (see \cite{SS}, \cite{2K3S})
\bean
&&F_{\pi^0\to\gamma e^+ e^-}(q_1^2)=F_{\pi^0\to\gamma^*\gamma^*}(q_1^2,0) \, \, .
\eean

      The wave function in (7) is a solution of a covariant two-body
    equation \cite{2K3S}, \cite{SKSOL}, \cite{FA}   that has a three
    dimensional form due to the
    use of a covariant single-time  method of describing of a relative
    motion of quarks in a system where $\pi$-meson has the 4-momentum P.
    It should be mentioned that the relativistic wave function is connected 
    with the vertex function  ${\Gamma(P|{\bf k}_1,{\bf k}_2)}$  
    according to formula

\bean
\Psi_{BP}({\bf k}_1)=\frac{\Gamma(P|{\bf k}_1,{\bf k}_2)}
{(k_1^2-m^2)(k_2^2-m^2)}
\eean
\noindent
    (we take the masses of quark and antiquark to be equal to $m_q$).
    The vertex function is used in diagram technique to include the 
    interactions between particles,  in our case quarks, which
    lines in Feynman diagrams  do  enter this vertex. 
    Formula (10) allows to establish a more close analogy with the 
    Feynman diagram approach  and with a quark triangle diagram often
    used to describe the decay process.

    Let us note that in perturbative leading order the vertex function,
    or  the  triangle Feynman 
    matrix  element (corresponding to decay amplitude),  is
    set  to be   $\Gamma(P|{\bf k}_1,{\bf k}_2)$=1. It should be 
    mentioned  also that the three-dimensional nature of the integration 
    in (7) over the 3-vector ${\bf k}_1$ moment components (as well
    as a  three-dimensial form of the wave function in (8)) is caused
    by a passing to a single-time  formalism \cite{SALP}, \cite{LTK}
    and by the  fact that  in difference  with the ordinary Feynman 
    diagram technique,  where the virtual particle  moments are 
    ``off the mass shell'', in 3-dimensional approach the momenta of 
    particles  are on a mass shell, i.e. $p^2=m^2$, but the equations
    for the bound state wave function are written ``off the energy 
    shell'' like  in ``old-fashioned'' perturbation theory.

    The interaction amplitude  $T^{\sigma_1\sigma_2}_{q\bar q\to\gamma^{*}V^{*}}
    (q_1;k_1|P)$  has the standard QFT form:
\bean
T^{\sigma_1\sigma_2}_{q\bar{q}\to\gamma^{*}V^{*}}(q_1,q_2;k_1|P)=
\nonumber
\eean
\bean
=\frac{4\pi\sqrt{\alpha} g s_q \bar{u}^{\sigma_2}_{\bar q}(k_2)\hat e_1
(\hat k_1-\hat q_1+m_q)\hat e_2(V-A)u^{\sigma_1}_q(k_1)}{(k_1-q_1)^2-m^2_q}
+(q_1\leftrightarrow q_2) \, \, .
\eean

 
    Here  $\hat e_2\equiv \gamma^\nu e_{2\nu}(q_2)$ with $e_{2\nu}(q_2)$, 
    being the polarization 4-vector of a photon (that would be finally
    treated as a real one) with  the 4-momentum $q_2$, while
    $\hat e_1\equiv \gamma^\alpha e_{1\alpha}(q_1)$
    with $e_{1\alpha}(q_1)$ being a  polarization 4-vector 
    of  a virtual boson (photon  or $W^{\pm}$) with the 4-momentum $q_1$. 
    The value  $s_{q}=\sqrt{n_c}\cdot \sum e_q^2$ includes a number of colors
    $n_c$ and the  summation is done over the squared charges of quarks
    appearing in a fermion loop of a diagram shown in Fig.1. Factor  $\alpha$ 
    is the electromagnetic  coupling constant while $g$ is equal to
    $\sqrt{\alpha}$ in a case of $\pi^0 \to \gamma+e^+e^-$ decay and
    $g=(eV_{ud})/2\sqrt{2}sin({\theta}_w)$ in a case of the process
    $\pi \to \gamma +e \nu$.

      Factor $(V-A)$ is defined by (6) and it takes into account
    the  structure of the vertex (see Fig.1) corresponding to an
    intermdediate  $V^*$ (= $\gamma^*/W^\pm$) boson coupling to 
    quarks.


      The spin structure of $\pi$-meson wave function is taken according 
      to  \cite{2K3S}, \cite{SS} as follows: 
\bean
\Psi_{\sigma_1\sigma_2BP}({\bf k}_1)=\bar{u}^{\sigma_2}_{\bar q}(k_2)\gamma^5 u^{\sigma_1}_q(k_1)
\frac{\tilde\phi_{BP}({\bf k}_1)}{2P\cdot k_1/M_\pi},
\eean

%

\noindent
    where $\sigma_1$ and $\sigma_2$ are quark  polarizations  and 
    $\tilde\phi_{BP}({\bf k}_1)$ is taken to be a  scalar function 
    because in what follows we shall consider $q\bar{q}$  $s$-state 
    (i.e. with zero orbital angular momentum $l=0$).
    It should be mentioned that if one shall put (12) (with setting
    $\frac{\tilde\phi_{BP}({\bf k}_1)}{2P\cdot k_1/M_\pi}=1$) into (7)
    and then substitute the wave function by the vertex function 
    according  to  formula (10) and take there 
    ${\Gamma(P|{\bf k}_1,{\bf k}_2)}$=1, then   an exact expresion of 
    QFT Feynamn matrix element would appear under the sign of the 
    integral. The nature
    of a 3-dimensial form of the integration can be easily understood on 
    the basis of widely used rather straitforward way of passing to a 
    3-dimensial formalism. In this approach one starts with the 
    expression of a decay amplitude taken as an integral convolution 
    ( with the  4-dimensional integration volume  element) 
    of a two-time 4-dimensional Bethe-Salpeter wave function 
    with a 4-dimensional Feynman amlitude. Then by  performing the
    subsequent  equating of fermion and antifermion individual  times 
    in Bethe-Salpeter  wave function by intoduction of $\delta$-function,
    having the difference of these 2 times as its argument
    (see  \cite{SALP}-- \cite{BST})
    one gets a 3-dimensional equation in the momentum space.

 
\section{Dip-effect in $\pi^0\to\gamma+e^+e^-$ decay formfactor.}


    In a case of  $\pi^0 \to \gamma+ee$ only the V=1 term in (11) is 
    taken in the amplitude, so it has a  pure quantum electrodynamical
    QED form.

     After substituting of such a final state interaction amplitude
    into (7) one has to perform the summation over spin polarizations,
    what leads to the appearence of the corresponding trace of
    $\gamma$ matrices, including those that were summed up with two 
    polarization 4-vectors $e_{1\mu}(q_1)$ and $e_{2\nu}(q_2)$ in (11).
    The Lorentz indexes of these $\gamma^\mu$ and   $\gamma^\nu$  
    matrices, associated with two fermion-boson interaction 
    vertices shown in Fig.1, define the Lorentz index structure of the 
    expresion for the calculated trace which is equal to $V_{\mu\nu}(q_1|P)$
    in (3). 
 
    Two polarization 4-vectors of photons  $e_{1\mu}(q_1)$ and 
    $e_{2\nu}(q_2)$  are not included, according to
    the definition (8), into the expression for the decay formfactor.
    The last one, thus, is a Lorentz scaler and is defined  only through 
    the quark block of the diagram shown in Fig.1, that includes only 
    two quark wave function  and quark variables which enter 
    the amplitude (11).



    After calculation of the trace and the  separation of its convolution with 
    photon polarization vectors from the ampitude  one may pass to performing
    the  integration over the angular   variables. 
    Those are left, in a case of $s$-state,
    only in the  denominator of quark  propagator in (11)

\bean
\frac{1}{(k_1-q_1)^2-m^2_q} = \frac{1}{2q\cdot k_1} \cdot 
\frac{1}{A+z},
\eean
where 
\bean
A=\frac{\ds q_1^2-2q_1^0k_1^0}{\ds 2q\cdot k_1}.
\eean

    Here $z=cos(\theta)=(\vec{q}_{1}\cdot \vec{k}_{1})/q k_{1}$ and the
   notations $k_{1}=|\vec{k}_{1}|$ ;  $q=|\vec{q}_{1}|$ are used.
   The following relations 
   are valid also: for a real photon momentum we have
   $q^2_2=q_{20}^2-(\vec{q_2})^2=0$, i.e. $q_{2}^{0}=|\vec{q}_{2}|$;
   for  quark 4-momentum  $k_{1}^2=m^2$, because, as it was mentioned before,
   within the approach used  in \cite{2K3S}  the 4-moments of particles are
   ``on the mass shell'' but out of the covariantly defined  "energy shell"
   \cite{FA}.

    We denote the square of the 4-momentum
    $q_{1}$=($q_{1}^0$, $\vec{q}_{1}$) of a virtual vector boson $V$ that 
    produce a final state lepton-antilepton pair  as $q_{1}^2=Q^2$,
    keeping for the  modulus $|\vec{q}_{1}|$ a notation $q=|\vec{q}_{1}|$.
    Thus, $A=(\ds Q^2-2q_1^0 k_1^0)/( 2q\cdot k_1)$
    and we get an expression 
\bean
\hspace*{-3mm} F_{\pi^0\to\gamma^*\gamma}(q^2_1)=\frac{8m_q s_q \alpha}
{\sqrt{2\pi}M_{\pi}}
\left\{2\pi\int\limits^{\infty}_0\frac{dk_1\cdot k_1^2}{2k_1^0}\cdot
\frac{\tilde\phi_{BP}({\bf k}_1)}{2q\cdot k_1}
\cdot \int_{-1}^{+1}\frac{dz}{\frac{\ds q_1^2-2q_1^0k_1^0}
{\ds 2q\cdot k_1}+z}  
\right\}.
\eean

      The appearance of the factor  $q=|\vec{q}_{1}|$ in the denominator
    of (13) and finally in (15) has an important  sequence that,
    possibly, may be experimentally observed.
     Really, due to the  relation  $q_{2}^2=(P-q_{1})^2$, following 
     from the 4-momentum conservation law $P= q_{1} + q_{2}$ we get a
     relation $M_{\pi}^2 - 2P q_{1} + q_{1}^2 =0$. This invariant formula
     can be rewritten as
\bean
        2P q_{1}=M_{\pi}^2 (1+x),\quad   x=Q^2/(M_\pi)^2,
\eean
    where from one can get in the pion rest frame ($\vec{P}=0$) the
    relations for the
    components of the 4-vector $q_{1}=(q_{1}^0$, $\vec{q}_{1})$ (keeping in
    mind our notation  $q=|\vec{q}_{1}|$ and the definition
    $q^2 \equiv |\vec{q}_{1}|^2 =q_{10}^2-(\vec{q_1})^2$):
\bean
 q_{1}^0 \equiv \frac{M_\pi^2 + Q^2}{2M_\pi} = M_{\pi} (1+x)/2,
\eean
\bean                                                            
  q \equiv |\vec{q}_{1}|  = \frac{M_\pi^2 - Q^2}{2M_\pi} = M_{\pi} (1-x)/2. 
\nonumber
\eean

     Thus we see that after integration of the propagator in (15) over
     $z=cos\theta$ one gets the prediction \cite{BERSNEL},\cite{2K3S}
\bean
     F(q_{1}^2) = F(Q^2) = F(x) \sim (1-x)^{-1}.                
\eean

     So, a possible growth of $F(x)$ in the region of $x\sim 1$, if
     it can be  observed  in a data, may serve as a confirmation that 
     the  choice of the propagater in (11), as well as of the quark
     amplitude  as a whole in a form of (11), i.e. in a standard for 
     perturbative QFT form, may be quite a reasonable one  and a
     consistent with data.

      A specific theoretical prediction of \cite{2K3S}, that
     is also connected with the  the form  of the propagator (13)
     from (11)( but is not comletely defined by it only)
     is about the  formfactor behavior  at small
     values of $x$ .
    
     Firstly let us mention that the investigation performed 
     in \cite{BERSNEL}, where the so called ``static''
     approximation for the wave function was used  ( what is in
     fact equivalent to ignoring of the effects of quarks motion 
     inside the meson) have shown that in this highly nonrelativistic 
     approximation the slope of the formfactor has a positive value.

     In  \cite{2K3S} another limiting case of ultrarelativistic
     quark  motion inside pion (i.e. when to the integral (15)
     large values of $k_1$  contribute mainly) was considerd.
     It was shown that in this  limit the derivative of the
     formfactor has a negative value  at sufficiently small $x$.
   
     Combining this observation, based on analytical calculations
     only, with the discussed above $\sim (1-x)^{-1}$ behaviour of
     the  formfactor at $x\sim 1$, i.e. where it has a positive 
     slope, one can suppose that for the relativistic bound state 
     systems (and a light ${\pi}$-meson is a good candidate in such
     a case) described with the relativistic wave functions,
     the formfactor $F(x)$ may have a minimum and, therefor, a 
     changing sign of its slope.(It is worth mentioning  that the
     present data for this slope, see, for instance, \cite{SACLPL},
     iclude the positive as well as negative values.) 

     The origin of such possible prediction can be  be understood
     from  the analysis of the structure of formulae (15) without 
     applying to any concrete form of the wave function.


%


      Really, the  integration over  $z$-variable  
     in  (15) leads to an  appearance  of a logarithmic function 
     \cite{BERSNEL}, \cite{2K3S}(which may  have different sign 
     in different regions of its argument) under the integral sign
\vskip 1mm
\bean
&&F_{\pi^0\to\gamma^*\gamma}(q^2_1)=\frac{8m_q s_q \alpha}
{\sqrt{2\pi}M_{\pi}}\cdot 
\left\{2\pi\int\limits^{\infty}_0\frac{dk_1\cdot k_1^2}{2k_1^0}\cdot
\frac{\tilde\phi_{BP}({\bf k}_1)}{2q_1\cdot k_1}
\cdot \ln{\left|\frac{q_1^2-2k_1^0q_1^0+2q\cdot k_1}
{q_1^2-2k_1^0q_1^0-2q\cdot k_1}\right|}
\right\}. 
\eean
\vskip 1mm

\noindent

       According to \cite{2K3S} the expression for the decay formfactor 
     $F_{\pi^0 \to \gamma+ee}(x)$ being normalized to the constant of 
     $\pi^0 \to 2 \gamma$  decay 

\vskip 5mm

\bean
&&F_{\pi^0\to\gamma^*\gamma}(q^2_1)=f_{\pi^0\to 2 \gamma}
\tilde F_{\pi^0\to\gamma^*\gamma}(q^2_1) \, \, .
\eean
\vskip 5mm

      may be transformed to  a form:
\bean
&&\tilde F_{\pi^0\to\gamma^*\gamma}(x)=\frac{1}{1-x}
\left\{1+\frac{1}{4}\frac{\int_0^{\infty}d\chi_k\phi(\chi_k)\ln{|X(x,\chi_k)|
} }{\int_0^{\infty}d\chi_k\phi(\chi_k)\chi_k}\right\} \, \, ,
\eean
\bean
&&X(x,\chi_k)=\frac{1-xe^{-\chi_k}(M_p/m_q-e^{-\chi_k})}{1-xe^{\chi_k}
(M_p/m_q-e^{\chi_k})} \, \, .
\eean
      where $4\pi\phi(\chi_k)=k_1 \tilde\phi_{BP}({\bf k}_1)$ with 
      $k_1=|\vec{k}_1|$.  The quark rapidity $\chi_k=ln[(k_1^0+k_1)/m]$ 
      corresponds to the following  parametrization of 4-vector components:

\bean
k_1^0=m_q ch \chi_k;  ~~~~~ k_1=|\vec{k}_1|=  m_q sh \chi_k.
\eean 
      The integral in the denominator of (21) defines the  
     ${\pi^0\to\gamma\gamma}$  decay constant  \cite{2S2},\cite{SS},
     \cite{BERSNEL}, \cite{2K3S}
\bean
f_{\pi^0\to 2 \gamma}=\frac{32(2\pi)^{3/2}m_q\alpha s_q}{M_p^2}
\cdot \int_0^{\infty}d\chi_k\phi(\chi_k)\chi_k \, \, .
\eean        

    It is clear from (22) that at $x=0$ one gets: $\tilde{F}(0)=1$ and
    thus
\bean
 F_{\pi^0 \to \gamma+ee}(0)=f_{\pi^0 \to 2\gamma}.
\eean                                                                  
      Formula (21)  presents the  formfactor normalized to unity  as 
     a product of two factors. It may  be also treated as a sum of
    two terms. One of them is $(1-x)^{-1}$. It defines the  monotonic
     growth of formfactor near $ x \sim 1$ without a changing of the
     sign of a curve slope. 

    The second one  contains (besides the integral over
    the wave function multiplied by log of (22)) also the factor
    $(1-x)$ in the denominator and serves as a small correction to the
    main  term $(1-x)^{-1}$. In \cite{2K3S} it was shown that  
    the sign of this correction depends on the sign of the 
    logarithmic function  $ln |X|$ under the integral sign
    and on what region of the  integration of this $ln |X|$ over
    the  rapidity dominats. 
    The last circumstance depends on a shape  of the wave function. 



    The wave function $\phi(\chi_k)$  takes into account 
    the bound state effect and, thus, has a nonperturbative nature.
    It serves  in intgral as a weight factor for  $ln |X|$ and
    defines what region of quark rapidity may give the most 
    contribution to the integral. Thus, the sign  of the logarithmic 
    function  $ln |X|$  in this region would define the sign of a
    small additional integral term (appearing here as a correction  
    to the leading term $(1-x)^{-1}$ in (21)) and thus it defines
    finally the sign of  the formfactor $F(x)$ slope in a region 
    of small values of $x$.

      In a case of $M_p/m_q \leq 1$  the numerator and
    the denominator in (22) are both positive, so the modulus sign
    in  $ln |X|$ can be omitted. Then it is  easy to check that for 
    the values of quark  rapidity  $\chi_{k}$,
    satisfying the relation $M_{\pi}\leq 2m_q ch(\chi_{k})$, 
    the  numerator in (22) would be  less than the  denominator and 
    thus   the  $ln |X|$ function would have a negative sign.
    So, we see that in a case when the binding energy of pion (as of 
    the $q\bar{q}$  sytem )is negative ( i.e. it may be  parametrized
    as follows:  $M_{\pi}=2m_q cos \beta$) 
    the condition  $M_{\pi}\leq 2m_q ch(\chi_{k})$
    would be satisfied and thus the  additional term to $(1-x)^{-1}$ 
    would be negative. It would lead, in principle, to a negative sign
    of the formfactor slope at small values of  $x$. The value of a 
    slope parameter depends on a shape of a particular wave function.

      From what was said above  it is clear that the logarithmic 
    function includes the information about perturbative amplitude
    (11).  In this sense the log function in (21)
    fulfils a job of a perturbative   probe by help
    of which one can get under the sign of the integral and  
    test the shape of the $q\bar{q}$ bound state wave function
    in momentum space by means of variation the of external kinematic 
    parameter $x=Q^{2}/M_{\pi}^2$ value.

    Now after these  general cosiderations it is a time to mention
    the results of \cite{2K3S} where for numerical calculation of
    the formfactor behaviour  two types of relativistic wave
    functions obtained in \cite{2K3S}  as the solutions of 2-body
    relativistic three-dimensional equations with QCD inspired 
    model potentials were used as well as one wave function that 
    is an exact solution of relativistic oscillator model.
    All of these model
    wave fuctions have lead to a negative sign of the formfactor
    slope at small values of $x$ (see below Fig.2 where the result
    obtained within the relativistic oscillator model is presented).
    The results of other two QCD models give curves with the position
    of their minumums being  from 20  to 40 percents  higher than that
    one of a curve shown in Fig.2.\\[-10pt]
\begin{figure}[h]
\hspace*{43mm} \includegraphics[width=8cm,height=9.5cm]{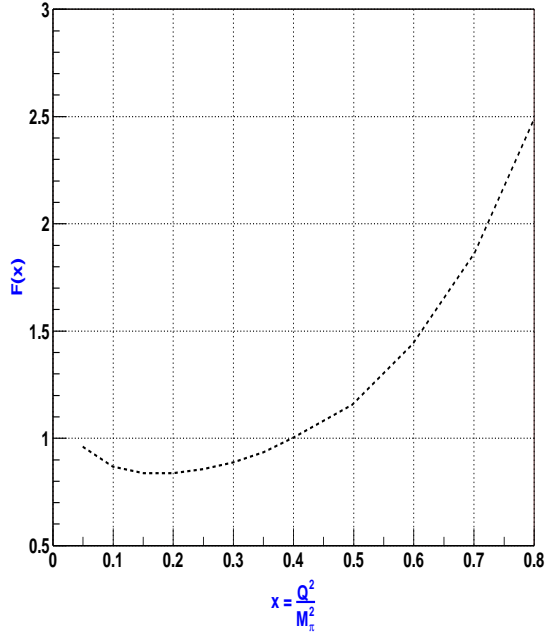}
\vskip-5mm
\caption{\hspace*{0.0cm} The title of figure 2.}
\end{figure}
                                                                           
%
        
%
%
%

     In this connection it is worth mentioning the result of 
   paper \cite{2K3SMU} where, for a sake of testing the method
   used in paper \cite{2K3S}, a decay of purely QED system of 
   muon-antimuon bound state into a real photon and $e^+e^-$
   pair was considered. The invariant mass of two bound muons
   is much more higher than the value of two electron masses.
   So there is a more wide interval of $Q^2$ values may be
   attainable as comparing with a case of $\pi$-meson decay.
 
   At the same time the well studied apparatus of QED is known
   to be experimentally  checked quite well and the tools for
   calculation of relativistic wave functions in a case of QED 
   interactions in purely electromagnetic systems are also
   existing. So a muon-antimuon bound state is a good place to
   test  the idea of the approach discussed above. 

     The output
   of this kind of work done in \cite{2K3SMU} was finding out
   that the same dip in the formfactor of
   $(\mu\bar{\mu}) \to\gamma + e^+e^-$ decay exists but its
   depth is of  about one order smaller than that one found in 
   $F_{\pi^0\to\gamma e^+ e^-}$ formfactor. This difference in
   a value of dip was obtained with the same expression for the
   amplitude (7) ( with $A=0$ ), but with the Coulomb-like
   relativistic wave function. 
   Thus, from here one may make a conclusion 
   that the depth of the formfactor dip is defined by a value
   of binding energy of a system, which was found to be higher
   in a case of  $\pi$ -meson within the theoretical models 
   considered in  \cite{2K3S}. 
   
     So, from comparison of the results of these two works, one
   may conclude that the value of the depth in dip, being measured
   at experiment, may give the information about the value of 
   the binding energy in $\pi$-meson as a $q\bar{q}$ bound state
   system.   


\section{Dip-effect in  $\pi^{\pm} \to \gamma + e^\pm \nu$  decay formfactor.}


     The detailed discussion of structure of the used expressions,
   perfomed in a previouse Section, allows now  to pass easily to a case 
   of  $\pi\to \gamma+e\nu$ processes.

      First let us note that  the factors 1)$\frac{1}{q_1^2}$; 2)$e$;
   3) $j_{V}^{\mu}(p_+,p_-)$  that follow each other in the expression (1)
   do   represent, respectively:

   1. the virtual photon propogator,
\bean
  {g^{\alpha\mu}}/{q_{1}^{2}} ,
\eean 

      It contains the metric tensor  ${g^{\alpha\mu}}$ which convolutes 
      the Lorentz index of the  $\gamma^{\alpha}$- matrix  (that was  
      included into  $\hat e_1\equiv \gamma^\alpha e_{1\alpha}(q_1)$ in 
      amplitude (11)) with the Lorentz index $\mu$ of lepton current 
      $j_{V}^{\mu}(p_+,p_-)$.
      This  $\gamma^{\alpha}$- matrix was splited from  
      the polarization 4-vector of a photon $e_{1\alpha}(q_1)$ while 
      extraction (according to formula (8)) of the formfactor
      $F_{\pi^0\to\gamma e^+ e^-}(q_1^2)$  from the amplitude (7):

   2. current $j_{V}^{\mu}(p_+,p_-)$, that describs the lepton-antilepton 
      pair production and is characterized by 

   3. external to current  $j_{V}^{\mu}(p_+,p_-)$ factor e, i.e.
      an electric charge of electron - the  QED coupling constant,
      not included into the expression for
      the  current  (5) for a sake of convention and for keeping the
      universal structure of current definition.

   All these three factors do correspond to the lines on the
   diagram of Fig.1 that are external to the formfactor, which one,
   as it was discussed previously, contains only $q\bar{q}$ wave 
   function and quark components of amplitude (11).

     Let us rewrite the formula (2) in an analogous way. For this aim
   we shall write the factors that have to appear according to the
   Feynman rules of Standard Model if we shall take them for the
   diagram shown at Fig.1:

   1.1 the virtual W-boson propogator  
\bean
\frac{[g^{\alpha\mu}-{( q_{1}^{\alpha}q_{1}^{\mu})}/{M_{W}^{2}}]}{({q_1}^{2} - M_{W}^{2})};
\eean         
    
   2.1 current $j^{\mu}_{V-A}(p_{\pm},p_{\nu})$ defined by (5) and (6),
   
   3.1 external to current $j^{\mu}_{V-A}(p_{\pm},p_{\nu}) $ factor of SM 
       coupling constants, defined for a case of $W$ exchange and the production
       of  a pair of electron and neutrino in a final state  $f=\gamma+e\nu$ 
       for $\pi^{\pm}$-decay as


\bean
e/(2\sqrt{2}sin(\theta_w)),  
\eean

       while for the vertex of diagram in Fig.1, where $W$ couples to quarks as
 
\bean
(eV_{ud})/(2\sqrt{2}sin(\theta_w)) . 
\eean                   
        
   The combination of all of this factors allows to write down the
 expression (2) in the following form 

\bean
M_{\pi^{\pm}\to\gamma e^{\pm}\nu}(q_1,q_2|P)&=&\left[F_{V}(q_1^2) \cdot V_
{\alpha\nu}(q_2|P)+F_A(q_1^2) \cdot A_{\alpha\nu}(q_2|P)\right]\times 
\nonumber
\eean
~\\[-25pt]
\bean
\times \left(e^2 \frac{V_{ud}}{8sin^2(\theta_w))}\right) \cdot
\frac{[g^{\alpha\mu}-{( q_{1}^{\alpha}q_{1}^{\mu})}/{M_{W}^{2}}]}{({q_1}^{2} - M_{W}^{2})}
 \cdot   j_{V-A}^{\mu}(p_{\pm},p_{\nu}) e^{\nu}(q_2) \, \, 
\eean

\noindent
which has  the sructure  analogous to formula (1) for QED process 
of  $\pi^0 \to \gamma + e^+e^-$ decay.

   If we shall consider in the last formula the limit $q_1^2 \ll  M_{W}^2$
   and take into  account the relation 
\bean
(e/(2\sqrt{2}sin(\theta_w))^2=M_{W}^2(G_{F}/\sqrt{2})  
\eean  
   then  we come to formula (2) that parametrizes  the amplitude of
 $\pi^{\pm} \to \gamma + e^\pm \nu$ process through the formfactors
 $F_{V}$ and  $F_{A}$. 

    Thus it is shown  that the formula (30) has the same QFT structure as 
  the formula (1), discussed in the previouse Section. Now if we shall 
  write down the analog of formula (8) for a case of  
  $\pi^{\pm} \to \gamma + e^\pm \nu$ decay and present the amplitude (7)
  as a sum of two terms according to two $V$ and $A$ parts of (V-A) factor
  in the amplitude (11), 
  then defining by help of polarization vectors  $e_1^{\mu}$ and $e_2^{\nu}$
  of virtual photon and the vitual $W$-boson   for each part the
  corresponding formfactor, we shall see that both of them would be defined
  by one and the same expression (21) as they are given by one and the
  same quark bloc like that one considered in a previuose Section.
   
    From here it is clear that the behaviour of the formfactors in a case
  of  $\pi^{\pm} \to \gamma + e^\pm \nu$ decay must have the same dip
  as shown in Fig.2 for   $\pi^{0} \to \gamma + e^{+}e{-}$ process.

\normalsize

\end{document}